\documentstyle[11pt,aaspp4]{article}

\newcommand{\fluxunit}{$ \ {\rm erg}\ {\rm cm}^{-2}\ {\rm s}^{-1} $}

\lefthead{Akiyama et al.}
\righthead{The hardest X-ray source in the {\it ASCA} LSS}

\begin{document}

\title{Optical Identification of the Hardest X-ray Source
in the {\it ASCA} Large Sky Survey}

\author{Masayuki Akiyama\altaffilmark{1,2,3} and Kouji Ohta\altaffilmark{1,2}}
\affil{Department of Astronomy, Kyoto University, Kyoto 606-8502, 
    Japan}

\author{Toru Yamada\altaffilmark{1}}
\affil{Astronomical Institute, Tohoku University, Sendai 980-8578, Japan}

\author{Michael Eracleous\altaffilmark{1,4}}
\affil{Department of Astronomy, University of California, Berkeley, CA 94720}

\author{Jules P. Halpern\altaffilmark{1}}
\affil{Department of Astronomy, Columbia University, New York, NY 10027}

\author{Nobunari Kashikawa and Masafumi Yagi} 
\affil{National Astronomical Observatory, Mitaka, Tokyo 181-8588, Japan}

\author{Wataru Kawasaki}
\affil{Department of Astronomy, University of Tokyo, Tokyo 113-8658, Japan}

\author{Masaaki Sakano\altaffilmark{3} and Takeshi Tsuru}
\affil{Department of Physics, Kyoto University, Kyoto 606-8502,
      Japan}

\and

\author{Yoshihiro Ueda and Tadayuki Takahashi}
\affil{Institute of Space and Astronautical Science,
Kanagawa 229-8510, Japan}
 
\altaffiltext{1}{Visiting Astronomer, Kitt Peak National Observatory, 
National Optical Astronomy Observatories, which is operated by the
Association of Universities for Research in Astronomy, Inc. (AURA)
under cooperative agreement with the National Science Foundation.} 
\altaffiltext{2}{Visiting Astronomer, University of Hawaii Observatory.}
\altaffiltext{3}{Research Fellow of the Japan Society for the Promotion 
of Science.}
\altaffiltext{4}{Hubble Fellow.}

\begin{abstract}
\noindent
We report the optical identification of the hardest X-ray source
(AX~J131501+3141) detected in an unbiased wide-area survey in the
0.5--10 keV band, the {\it ASCA} Large
Sky Survey (LSS).
The X-ray spectrum of the source is very hard
and is well reproduced by a power law 
component ($\Gamma = 1.5^{+0.7}_{-0.6}$) with $N_{\rm H} = 6^{+4}_{-2}
\times 10^{22}$ cm$^{-2}$ (\cite{sak98}).
We have found a galaxy with $R=15.62$~mag near the center of the 
error circle for the X-ray source.
The optical spectrum of the galaxy shows
only narrow emission lines whose ratios correspond to
those of a type 2 Seyfert galaxy at z = 0.072, implying an
absorption-corrected X-ray luminosity of  
$2 \times 10^{43}$~erg~sec$^{-1}$ (2--10 keV) and
$M_{\rm B} = -20.93$ mag. 
A radio point source is also associated with the center
of the galaxy.
We thus identify the X-ray source with this galaxy as an
obscured AGN.
The hidden nature of the nucleus of the 
galaxy in the optical band is consistent with the X-ray spectrum.
These results support the idea that the obscured AGNs/QSOs contribute
significantly to the cosmic X-ray background
in the hard band at the faint flux level. 

\end{abstract}


\section{Introduction}
\noindent
Many efforts have been made to understand the origin of 
the cosmic X-ray background (CXB).
Recently in the {\it ROSAT\/} deep surveys,
$\sim$ 60\% of the CXB in the 0.5--2~keV band has been resolved
into discrete sources (\cite{has93}). 
On the other hand, in the harder 2--10~keV band,
where the bulk of the energy density of the CXB resides, only $\sim 3$\%
of the CXB was resolved into discrete sources (\cite{pic82}).
In the soft X-ray band (0.5--2~keV)
type~1 AGNs are the main contributers to the CXB 
within a flux level of $ 1.6 \times 10^{-15} $ \fluxunit\ (\cite{mch98}).
However, they have X-ray power-law spectra
($\Gamma = 1.7$, \cite{tur89})
significantly 
softer than that of the CXB in the 2--10~keV band
($\Gamma = 1.4 \sim 1.5$, \cite{gen95}, \cite{ish96}).
Therefore, there must be 
objects which have harder X-ray spectra than type 1~AGNs
and contribute significantly to the CXB in this
band. 

To study the nature of X-ray sources in the hard band,
we are now conducting an unbiased large and deep
survey with the {\it ASCA} (\cite{tan94}) near the north Galactic pole.
This is the Large Sky Survey (hereafter LSS,
\cite{ino96}, \cite{ued96}, \cite{ued98}).
The flux limit of the LSS ($\sim$ 1 $\times 10^{-13}$ \fluxunit\
in the 2--10~keV band) is 100 times deeper than
the {\it HEAO1\/}~A2 survey, which was the deepest
survey in the hard band (\cite{pic82}).
We have surveyed 7.0 deg$^2$ and detected 43 sources
above the 4$\sigma$ level in the 2--10 keV band (\cite{ued96}),
corresponding to $\sim$ 30\% of 
the CXB in this band.
The average photon index of the X-ray sources detected
in the flux range between 1 $\times 10^{-13}$
and 5 $\times 10^{-13}$ \fluxunit\
is $\Gamma = 1.5 \pm 0.2$ (\cite{ued98}),  
which is significantly harder than the spectra of
X-ray sources detected in shallower surveys in the
2--10~keV band and close to that of the CXB.
Thus, we are now revealing the presence of 
the hard sources which are responsible
for most of the CXB energy density. Identification of these hard
sources is clearly an important next step.

In the LSS, we have discovered the very hard X-ray source
AX~J131501+3141 whose power-law index is by far the hardest among the LSS
sources.  This object provides a good opportunity
to examine the nature of the hard X-ray sources found in our
systematic survey, and the
result will be a key for the understanding of
the origin of the CXB in the hard band.
In this paper, we report on the optical identification and properties
of AX~J131501+3141, and their implications for the origins of the CXB.
Results of the X-ray observations are presented in detail in a
separate paper (\cite{sak98}).
Throughout this paper, we use $q_0=0.5$ and $H_0=50$ km s$^{-1}$ Mpc$^{-1}$.

\section{Identification of AX~J131501+3141}

\subsection{X-ray Properties}
\noindent
AX~J131501+3141 is the hardest X-ray source detected in the LSS.
The apparent X-ray spectrum from its initial observation
can be fitted by a single power-law model 
with a photon index $\Gamma = -0.18 \pm 0.31$
($f_{\rm E} \propto {\rm E}^{-(\Gamma-1)}$) in 0.7 -- 10 keV band.
Since the photon indices of the other X-ray sources detected in the LSS 
are distributed between 0.4 and 4.0
in the flux range 
from 1 $\times 10^{-13}$ to 1 $\times 10^{-12}$ \fluxunit\
(2--10~keV), 
the photon index of AX~J131501+3141
is outstandingly hard.

Subsequently deeper X-ray observations of this object
totaling 100,000~s were made
to learn more details of its X-ray properties (\cite{sak98}).
The results given in Sakano et al. (1998) are as follows.
The flux of AX~J131501+3141 in the 2--10~keV band is  
$\sim 5 \times 10^{-13}$ \fluxunit,
while in the 0.5 -- 2 keV band 
no X-ray flux is detected above a 3 sigma level
and the corresponding upper limit is $1.2 \times 10^{-14}$ \fluxunit.
The X-ray spectral shape seems to be a power law 
with a low energy cut-off at $\sim 2-3$~keV;
the best fitted parameters are 
$\Gamma = 1.5^{+0.7}_{-0.6}$ and
absorbing column density $N_{\rm H} = 6^{+4}_{-2}
\times 10^{22}$ cm$^{-2}$
in the observed frame.
The Galactic column density of neutral hydrogen in this direction is 
$N_{\rm HI} = 1.1 \times 10^{20}$ cm$^{-2}$ (\cite{sta92}).
The deep observations were made at two epochs (1995 December and
1996 June) and
30\% variability was detected.
From the deep X-ray observations,
the position of AX~J131501+3141 is accurately determined to be 
$\alpha$=13$^{\rm h}$15$^{\rm m}00^{\rm s}.9$, 
$\delta$=31$^{\circ}$41$'$28$''$ (J2000) with an error radius of
$30^{\prime\prime}$ (95\% confidence).

\subsection{Optical Imaging Observations of AX~J131501+3141}
\noindent
To identify the optical counterpart of AX~J131501+3141,
we made imaging observations
with a Tektronix 2048$\times$2048 CCD on 
the University of Hawaii 88$^{\prime\prime}$ telescope in 1996 April.
Images were obtained in two bands, $B$ and $R$, 
with a spatial sampling of $0.\!^{\prime\prime}22$~pix$^{-1}$.
The exposure times were 20 and 15 min for the $B$ and $R$ bands, 
respectively.
The FWHM of the seeing in the images was
$1.\!^{\prime\prime}1$ in $B$ and $1.\!^{\prime\prime}0$ in $R$.

\placefigure{fig1}

Figure \ref{fig1} shows the $R$ band image centered on the position of  
AX~J131501+3141 with its error circle of radius $30^{\prime\prime}$.
In the error circle, there are one bright galaxy and a few faint objects.
The coordinates of the bright galaxy are
$\alpha$=13$^{\rm h}$15$^{\rm m}$01$^{\rm s}.15$,
$\delta$=31$^{\circ}$41$'$28$''.1$ (J2000),
which is only $3.\!^{\prime\prime}2$ from the best position of the 
X-ray source. Hence this is the most promising candidate for the optical
counterpart. The galaxy has total magnitudes $R=15.62 \pm 0.02$
and $B=17.25 \pm 0.02$,
and is cataloged as an S0 galaxy with $B=17.2$ in Slezak et al. (1988).
Hereafter, we call this bright galaxy ``galaxy~A''.

The other objects in the error circle
are fainter than $B=22.4$.  Since
type~1 AGNs are the most X-ray loud objects among the
known extragalactic X-ray sources, we expect that the optical counterpart
of AX J131501+3141 must be brighter than $B=19.9$ on the basis of 
its 2--10~keV X-ray flux, assuming $\Gamma=1.7$
together with the X-ray-to-optical 
flux ratio of type~1 AGNs identified in the
Cambridge-Cambridge {\it ROSAT\/} Serendipity Survey (\cite{boy95}).
Since $B=22.4$ is more than one order of 
magnitude fainter, none of these other objects are likely
to be the optical counterpart.

The chance probability of one galaxy brighter than 16.0 mag falling
in the error circle of the X-ray source is only 8 $\times 10^{-3}$.
This is obtained
by adopting a surface density of such galaxies
of 38 deg$^{-2}$ (e.g., \cite{ste86}).
This low probability reinforces
the conclusion that the galaxy~A is the optical counterpart of AX~J131501+3141.

\subsection{A Radio Source in the Error Circle of AX~J131501+3141}
\noindent
In the error circle of AX~J131501+3141, a radio source is detected
in the FIRST survey.
The FIRST survey is a radio source survey conducted with the Very Large Array
in the 1.4 GHz band with a $5\sigma$
limiting flux of 1~mJy (\cite{bec95}).
The coordinates of the radio source are 
$\alpha$=13$^{\rm h}$15$^{\rm m}$01$^{\rm s}$.19, 
$\delta$=31$^{\circ}$41$'$29$''$.1 (J2000),
which is very close ($\sim 1^{\prime\prime}$) to the center
of galaxy~A (Table \ref{tab1}).
The radio source is point-like and no structure can be seen.
Considering the positional accuracy of the radio source
($\sim 1^{\prime\prime}$)
and that of the optical galaxy center ($0.\!^{\prime\prime}8$),
we regard the radio source as being associated with the center of
galaxy~A.

In the course of the optical follow-up program of the LSS,
we examined the distribution of FIRST radio sources
around LSS X-ray sources to evaluate the cross-correlation
between radio and X-ray sources, and found
a $3.5\sigma$ correlation within a radius of $0.\!^{\prime}6$
from the X-ray sources (\cite{aki98}).
This radius corresponds to the error radius of the 
X-ray source positions.
The presence of a significant 
correlation between the radio and X-ray sources implies that
a FIRST radio source within $0.\!^{\prime}6$ of an LSS X-ray source is 
the radio counterpart of the X-ray source with a probability
of more than 80\%.  This is an additional reason why
galaxy~A, which is also the radio source 
detected near the center of the error circle of AX~J131501+3141,
is very likely to be the optical counterpart of the X-ray source.

In summary, considering all the results discussed in \S 2, we
conclude that galaxy~A and its central radio source are
to be identified with AX~J131501+3141.
In the next section, we will
present further evidence on the nature of galaxy~A.

\placetable{tab1}

\section{Nature of AX~J131501+3141}

\subsection{Type 2 Seyfert at z=0.072}
\noindent
We made a spectroscopic observation of AX~J131501+3141 (galaxy~A) 
with the multi-slit spectrograph (CryoCam)
on the KPNO Mayall 4m telescope 
on 1996 April 9.
We used grism 730 and covered the
spectral range from 6000 \AA\ to 9000 \AA\ 
with a resolution of 22 \AA.
We used a $2.\!^{\prime\prime}5$
wide slit. The spectral sampling was 4.3 \AA\ pix$^{-1}$ and
the spatial resolution was $0.\!^{\prime\prime}84$~pix$^{-1}$.
The FWHM of the seeing was $\sim 1^{\prime\prime}$.
The exposure time was 30 min.
The spectrum was summed within a $4^{\prime\prime}$
width along the slit direction
(P.A. $62.\!^{\circ}5$) centered on the nucleus of the galaxy.

We detected emission lines of 
H$\alpha$, [N~II]$\lambda\lambda$6548,6583, and
[SII]$\lambda\lambda$6717,6731 clearly but no broad H$\alpha$ emission line.
The logarithmic line ratios are log([N~II]$\lambda$6583/H$\alpha$)=0.28 and
log([S~II]$\lambda$6717+6731/H$\alpha$)=$-0.07$,
which imply that this object is classified as a
type~2 Seyfert or LINER (\cite{vei87}).
To be more definitive in the classification,
spectrum in the blue wavelength region is needed.
The redshift of this object turned out to be $z = 0.072$ and 
the resulting luminosities are
$L_{\rm X}$ = 2 $\times 10^{43}$ ergs~s$^{-1}$ in the 2--10~keV band 
after absorption-correction and $M_{\rm B} = -20.93$ mag
without K-correction. 
The power of the nuclear activity in the hard X-ray band 
is as large as that of a typical
Seyfert galaxy (the knee of the luminosity function of 
type~1 AGNs is $L_{\rm x} \sim 3 \times 10^{43}$ erg~s$^{-1}$ 
in the 0.5--3.5~keV band according to \cite{jon97}).

In order to obtain a spectrum in a blue region, 
we made another spectroscopic 
observation of galaxy~A with the KPNO 2.1m telescope and 
the Goldcam spectrograph on 1997 June 8.
Three 30 min exposures were taken through a $1.\!^{\prime\prime}9$ slit at
P.A.= 90$^{\circ}$. The spectra covered the
wavelength range 3866--7511\AA\ with a spectral resolution of 5.1\AA.
The uncertainty in the wavelength scale is 0.1 \AA.
The flux from the nuclear region dominates the optical 
spectrum and it is only
slightly resolved spatially in this observation.  Therefore, we summed the
spectrum using an optimal extraction method applied to a
$13^{\prime\prime}$ window along the slit.
Atmospheric absorption band correction
and flux calibration were then carried out with the help of spectra of 
standard stars.
The resulting spectrum is shown in Figure \ref{fig2}.
We can clearly see emission lines of [O~II]$\lambda$3727, H$\beta$,
[O~III]$\lambda\lambda$4959,5007,
H$\alpha$, [N~II]$\lambda\lambda$6548,6583, and
[SII]$\lambda\lambda$6717,6731.

\placefigure{fig2}

The continuum contains a great deal of starlight, as evidenced by the
Ca~II H \& K, H$\gamma$, Mg~I~b, and Na~I~D absorption lines.
Also, the H$\beta$ line shows an absorption plus emission structure.
In order to obtain the more accurate measurements of
the weak emission lines, we must subtract the stellar continuum spectrum.
The model spectrum consists 
of an elliptical galaxy stellar template (NGC~5332) and a featureless
power-law continuum, with the relative strengths of the
two components adjusted to provide the smoothest fit to the continuum
in the neighborhood of the emission lines.  The spectrum obtained
is also shown in Figure \ref{fig2}. 
Using the subtracted spectrum (Figure \ref{fig2}b dashed line),
fluxes of the emission lines can be measured more accurately,
though there are still
obvious residuals due to the
mismatch between the stellar populations of
the two galaxies.

\placetable{tab2}

The resulting logarithmic line ratios turn out to be
log([N~II]$\lambda$6583/H$\alpha$) = 0.06,
log([S~II]$\lambda$6717+6731/H$\alpha$) = $-0.31$,
and log([O~III]$\lambda$5007/H$\beta$) = 0.65
(see Table \ref{tab2}).
The ratio of [O~III]$\lambda$5007/H$\beta$ implies that the galaxy~A
is a Seyfert rather than LINER (\cite{vei87}).
These emission lines were resolved in the spectrum, with widths
all approximately the same and consistent with 
FWHM = $325 \pm 25$ km~s$^{-1}$.
There is no hint of the presence of a broad-line component of H$\alpha$ 
or H$\beta$, even after subtraction of the continuum.
Therefore we conclude galaxy~A is classified as a type~2 Seyfert.

One interesting feature in the spectrum is that
the intensity of [O~II]$\lambda$3727 emission line is
strong in galaxy~A.
In the [O~II]$\lambda$3727/[O~III]$\lambda$5007 vs.
[O~III]$\lambda$5007/H$\beta$ plane
(\cite{fer83}), galaxy~A is located between typical type~2 Seyferts and
LINERs  
due to its
large log([O~II]$\lambda$3727/[O~III]$\lambda$5007) values (0.004).
After correcting for reddening using the large Balmer decrement
(H$\alpha$/H$\beta$ = 7.58), galaxy~A 
falls among typical LINERs in this diagram because of the resulting
larger value of log([O~II]$\lambda$3727/[O~III]$\lambda$5007) = 0.36.
There are two possible explanations for the ratio of the oxygen emission
lines:
1) a small ionization parameter (U $\sim 10^{-3}$) 
in the nucleus, similar to LINERs, and 2)
contamination by starforming regions in the disk of the galaxy,
which would enhance [O~II]$\lambda$3727 relative to [O~III]$\lambda$5007.
We favor the second explanation because a LINER classification
is not supported by the other defining line ratio,
[O~I]$\lambda$6300/[O~III]$\lambda$5007,
 which is less than 1/3 in this spectrum,
and also because the distribution of the
H$\alpha$ emission along the slit in
the spectra taken with the 2.1m telescope is 
slightly more extended than that of [N~II]$\lambda$6583.
The spatial extent of the H$\alpha$ line is 4.2~pix FWHM
($3.\!^{\prime\prime}3$), while
that of [N~II]$\lambda$6583 is 3.7~pix ($2.\!^{\prime\prime}9$).
Unfortunately
we cannot separate the nuclear component from an extended component
in [O~II]$\lambda$3727 and H$\beta$ 
the signal-to-noise ratios 
of the lines are low
and since the focus of the spectrum at the blue end is poor.
The difference between H$\alpha$ and [N~II]$\lambda$6583 region is 
also confirmed in the spectrum taken with the 4m telescope.
In addition, the color of the disk of galaxy~A is $B-R = 1.3$, which
is bluer than the bulge of the galaxy.
These results suggest there is some contamination to the
nuclear emission lines by 
star formation in the disk of the galaxy.

\subsection{Hidden Nucleus of the Galaxy}
\noindent
The optical image (Figure \ref{fig1}) of AX~J131501+3141 shows
no point-like structure at the 
center of the galaxy, suggesting that
the majority of 
the optical continuum photons in the central 
region of the galaxy originate in stars in the bulge.
The column density to the nuclear X-ray source,
$N_{\rm H} = 6 \times 10^{22}$~cm$^{-2}$ (\cite{sak98}),
corresponds to $A_V \approx 33$ mag according to the relations 
$N_{\rm HI+H_2}/E(B-V) = 5.8 \times 10^{21}$ cm$^{-2}$~mag$^{-1}$
(\cite{boh78}) and $A_V/E(B-V)$ = 3.14 (\cite{sea79}),
which is large enough to hide the optical nuclear source.
Thus the absence of a blue point-like nuclear source in the optical
light is consistent with the large column density derived from the X-ray
spectrum.

The large Balmer decrement (H$\alpha$/H$\beta$ = 7.58)
could be the result of extinction 
by a material distributed within or around the narrow-line region,
rather than on the scale of the galaxy (e.g., the disk of the edge-on galaxy).
If the large Balmer decrement is
the result of reddening, and we assume that the 
intrinsic ratio of H$\alpha$ to H$\beta$ is 3.1, the
color excess $E(B-V)_{\rm Balmer}$ is required to be 0.82.
This value corresponds to the column density to the narrow-line region of
$N_{\rm HI+H_2} = 4.8 \times 10^{21}$~cm$^{-2}$.
On the other hand, the $B-R$ color in a $5^{\prime\prime}$ aperture
centered on galaxy~A  
is $1.96 \pm 0.03$ mag, which is redder than those of 
normal elliptical galaxies
($B-R = 1.47 \pm 0.05$ mag, \cite{but94}, \cite{but95}).
If the intrinsic bulge color is the same as a typical color
of an elliptical galaxy, the color excess of the bulge  
$E(B-V)_{\rm bulge}$ is 0.28, if we adopt the relation
$E(B-V) = E(B-R)/1.76$
(\cite{sea79}).
This value is smaller than that to the narrow-line region.
Thus the difference between $E(B-V)_{\rm Balmer}$ and 
$E(B-V)_{\rm bulge}$ suggests that dust could be distributed
on the scale of the narrow-line region, and smaller than
that of the bulge.

\subsection{Spectral Energy Distribution}

\placetable{tab3}
\placefigure{fig3}
\noindent
Using the photometric results from radio to the X-ray wavelength,
we can compare the spectral energy distribution (SED) of galaxy~A
with other type~2 Seyferts.
Photometric results on the galaxy are summarized in Table \ref{tab3}.
[In the infrared we use an upper limit from 
the {\it IRAS} Faint Source Catalog (\cite{mos92}) in
the region containing galaxy~A].
Figure \ref{fig3} shows the SED of
galaxy~A as well as those of type~2 Seyferts taken from 
Mas-Hesse et al. (1995). The SEDs are normalized
such that they have the same flux density as galaxy~A at 
the optical wavelength
($\sim$ {\it B} band).
Since the logarithmic flux density ratio between radio (1.4 GHz) and 
optical ($B$ band) is $-4.83$, 
which is similar to the other radio-quiet type~2 Seyferts,
galaxy~A is a radio-quiet object.
The flux density in the hard X-ray band is located within 
a scatter of other type 2 Seyferts,
galaxy~A has a typical value of the flux density ratio of 
the hard X-ray to optical region as type~2 Seyferts.

\subsection{An Interacting Galaxy?}

\placefigure{fig4}
\noindent
In Figure \ref{fig4}, we show a wide image of
of AX~J131501+3141.
Around AX~J131501+3141, we can see two other bright galaxies
of similar size, one (galaxy~B) is located at $67^{\prime\prime}$ east,
and the other (galaxy~C) is located at $55^{\prime\prime}$ south-southeast
of AX~J131501+3141.
In course of the multi-slit spectroscopy,
we also observed these galaxies.
The results are shown in Table \ref{tab2}.
Since galaxy~C has $z=0.189$, its closeness 
to galaxy~A is just a chance
projection.
Galaxy~B has almost the same redshift (0.072) as galaxy~A. 
The velocity
difference between galaxy~A and galaxy~B is 90 km~s$^{-1}$,
and their projected separation is $67^{\prime\prime}$,
which corresponds to 124~kpc at $z=0.072$.
The magnitudes of galaxy B are $B=18.08$ and $R=16.90$, and
its morphology is that of a normal spiral galaxy without 
any clear distortion.
However, the spectrum of galaxy~B
has emission lines which are
H~II-region like (\cite{vei87}).
The $B-R$ color of its disk is 1.0 mag 
which is bluer than the disk of galaxy~A.
Thus these results would indicate enhanced star-formation in the disk 
of galaxy~B.
There is a possibility that the interaction
is a driver of both the nuclear activity in galaxy~A,
and the enhanced star-formation in the disk of the galaxy~B.

\section{Implications for the Nature of Hard X-ray Sources}
\noindent
The hardest X-ray source detected in the continuous 
7.0~deg$^{2}$ area of the LSS has been identified
with a type~2 Seyfert 
having the column density
to the nucleus of 
$N_{\rm H} = 10^{22.8}$~cm$^{-2}$ in the local universe (z$\le$0.1).
Surface density of such object is consistent with recent models
of the origin of the CXB by 
e.g., Madau et al. (1994) and Comastri et al. (1995) as described below.
They propose that 
type~2 AGNs are the hard X-ray sources 
which need to exist to account for the hardness of the spectrum of the CXB,
because type~2 AGNs have harder X-ray spectra than type~1 AGNs 
due to the absorption of their soft X-ray photons.
In the surveyed area,
we have estimated the number density of type~2 AGNs
which have the column density
of $N_{\rm H}=10^{22-23}$~cm$^{-2}$
and X-ray flux larger than
$5 \times 10^{-13}$ \fluxunit\ in the 2--10~keV band
i.e., the flux of AX~J131501+3141. 
According to the model of Comastri et al. (1995),
the expected number is 1.2.
Observed surface number density is one (AX~J131501+3141),
if there are no more such objects in the LSS region.
In fact, we cannot expect the presence of more such objects
because of the following two reasons.
1) The AGNs with $N_{\rm H}=10^{22-23}$~cm$^{-2}$
in the local universe should have very hard X-ray spectra as like
the hardest source; but no such hard source is found in the LSS region
above $5 \times 10^{-13}$ \fluxunit\
(\cite{ued96}). 
2) High redshift AGNs with $N_{\rm H}=10^{22-23}$~cm$^{-2}$ have
softer X-ray spectra than the hardest source in the observed frame
due to K-correction and may exist in the LSS region.
However the probability that such high-z object exists is very low.
Because the predicted redshift distribution of absorbed AGNs above
$5 \times 10^{-13}$ \fluxunit (2--10~keV)
is peaked in z$\le$0.1 and declining with z in z$\ge$0.1, if we
adopt the evolution of X-ray luminosity 
function for type~1 AGNs (\cite{jon97}) and assume the number ratio
of type~1 to type~2 AGNs is constant with z.

In other surveys some hard X-ray sources have been also identified with
absorbed AGNs. In Iwasawa et al. (1997), 
a hard X-ray source (AX~J1749+684) 
detected serendipitously was identified with a
nearby ($z=0.05$) obscured AGN.
The hard X-ray flux of the object is 9.6 $\times 10^{-13}$ \fluxunit,
2 times brighter than AX~J131501+3141.
A fainter hard X-ray source (AX~J08494+4454) with
$f_x = 9.0 \times 10^{-14}$ \fluxunit in the 2--10~keV band,
was identified with a type~2 QSO at $z=0.9$
in course of optical follow-up observations of the {\it ASCA} Lynx
deep survey (\cite{oht96}).
A common feature of these three type~2 AGNs is
that they are hard X-ray sources with radio detections.
The $\alpha_{rx}$ ($f_{\nu} \propto \nu^{-\alpha}$)
values of these three objects are 
0.67 (AX~J131501+3141), 0.67 (AX~J1749+684) and 0.60 (AX~J08494+4454).
The agreement of these indices suggests that a hard X-ray survey
together with a radio survey is a fairly efficient way to select 
obscured type~2 AGNs.

Recently, in some optical follow-up programs for the {\it ROSAT\/} deep fields,
faint X-ray sources have been identified with narrow 
emission-line galaxies (NELGs) 
such as type~2 Seyferts, star-forming galaxies, and LINERs
(\cite{boy95b}, \cite{gri96}, \cite{mch98}).
These objects are thought to be possible candidates for hard sources,
but their X-ray spectra in the hard band have not been established.
With respect to their soft X-rays,
their 0.5--2~keV to optical flux ratios are 
one order of magnitude larger than those of
normal galaxies and one order of magnitude smaller
than those of QSOs (\cite{gri96}).
If we use the upper-limit of AX~J131501+3141 in the 0.5--2 keV band,
we find that its X-ray to optical flux ratio is
more than 10 times smaller than those of 
typical NELGs, and closer to the normal galaxies
($\log[f_{x}(0.5-2 keV)/f_{\rm B}]=-1.52$).
Thus AX~J131501+3141 may be different from
the NELGs detected in the {\it ROSAT} surveys.
However, it is possible that we are seeing the same population of
objects at lower redshifts.
To establish the nature of the hard X-ray sources,
identification of all the X-ray sources
detected in the LSS is clearly necessary.

\acknowledgments{
MA, KO, and TY would like to thank S.Okamura, M.Sekiguchi and 
the MOSAIC CCD camera team, and staff members of the KISO observatory for 
their support during the imaging observations.
MA thanks K.Aoki and T.T.Takeuchi 
for their help.
KO is grateful to the hospitality during his stay at the IfA, UH,
where a part of this work was done.
This research has made use of NASA/IPAC Extragalactic Database (NED),
which is operated by the Jet Propulsion Laboratory, Caltech,
under contract with the National Aeronautics and Space Administration. 
MA and MS acknowledge support from an 
Research Fellowships of the Japan Society for the Promotion of Science
for Young Scientists.
The optical follow-up program is supported by grand-in-aids from the 
Ministory of Education, Science, Sports and Culture of Japan
(06640351, 08740171, 09740173) and from the Sumitomo Foundation.
}

\clearpage

\figcaption[]{$R$ band image 
with the error circle of $30^{\prime\prime}$ radius
centered on the X-ray position of AX~J131501+3141.
North is up and east is left.
 \label{fig1}}

\figcaption[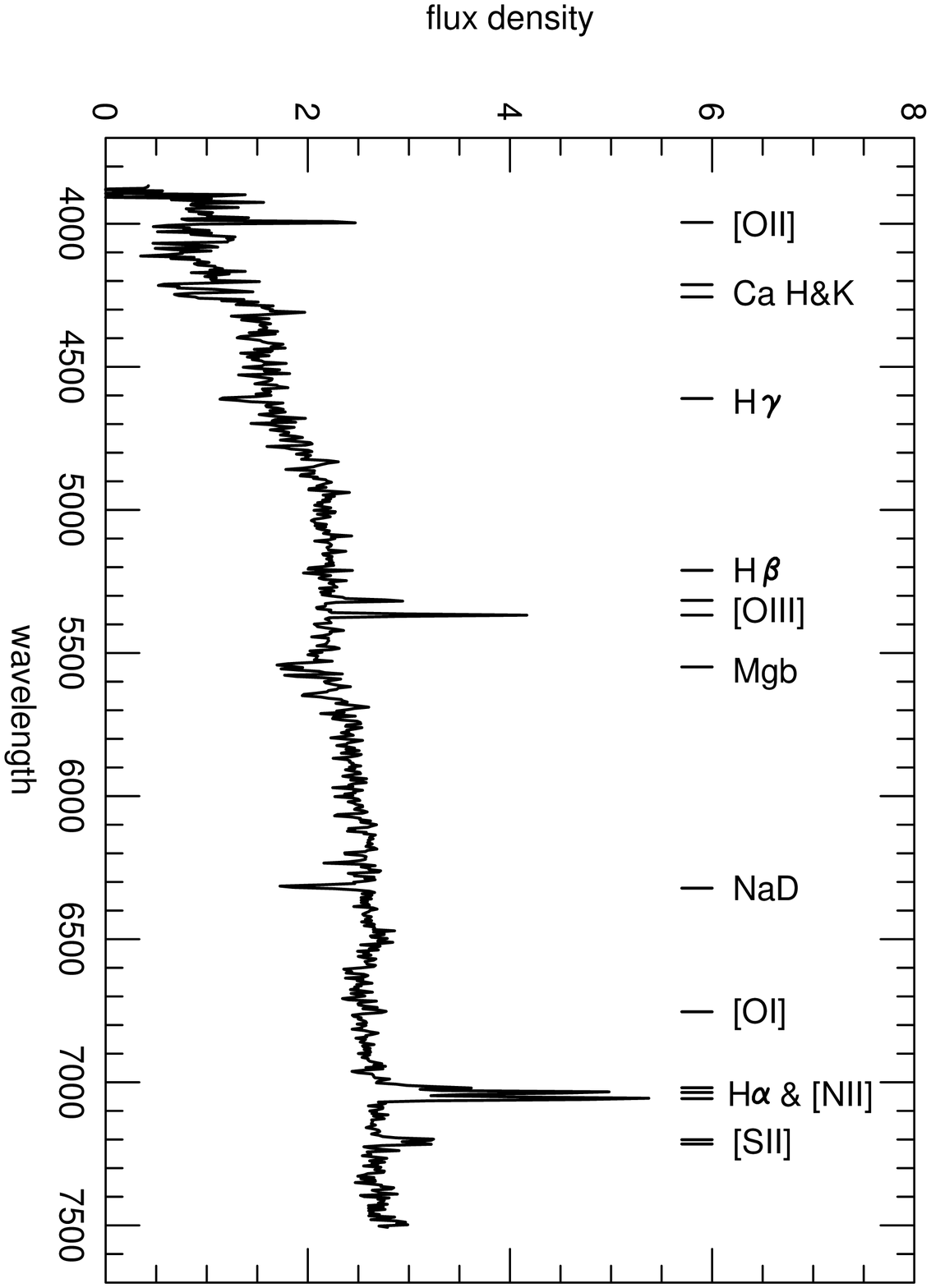, 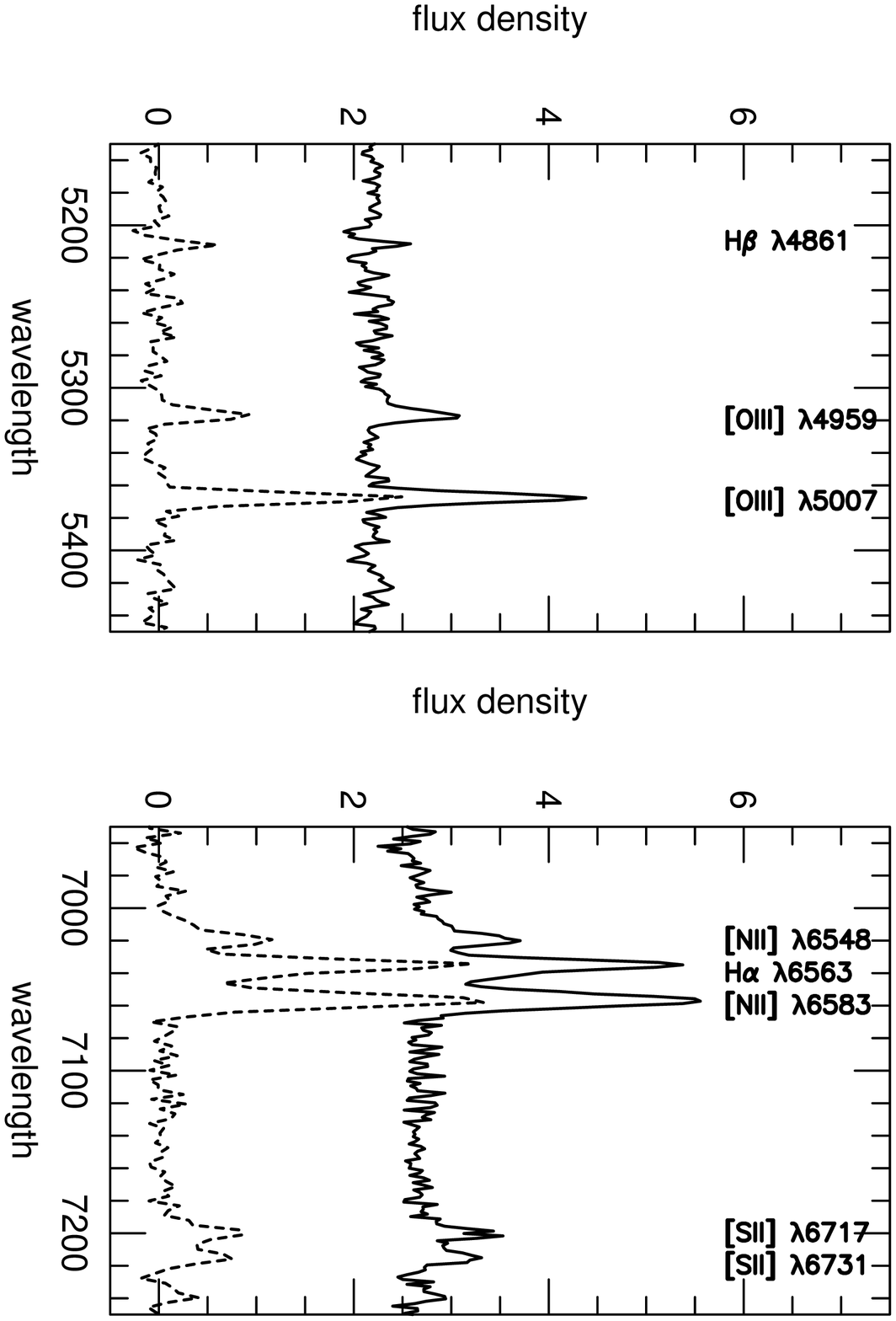]{
a) Optical spectrum of the nuclear region of galaxy~A.
The abscissa shows observed wavelength in \AA\ and the ordinates
shows flux density in $10^{-16}$ erg s$^{-1}$ cm$^{-2}$ \AA$^{-1}$.
Identified emission and absorption lines are marked at the top of the figure.
b) Close up view of the optical spectra. 
The abscissa and ordinates are same as in a).
Left panel shows H$\beta$ region,
 right panel shows H$\alpha$ region.
Dashed line in each panel shows the spectra after subtracting 
of the model stellar continuum (see text). 
 \label{fig2}}

\figcaption[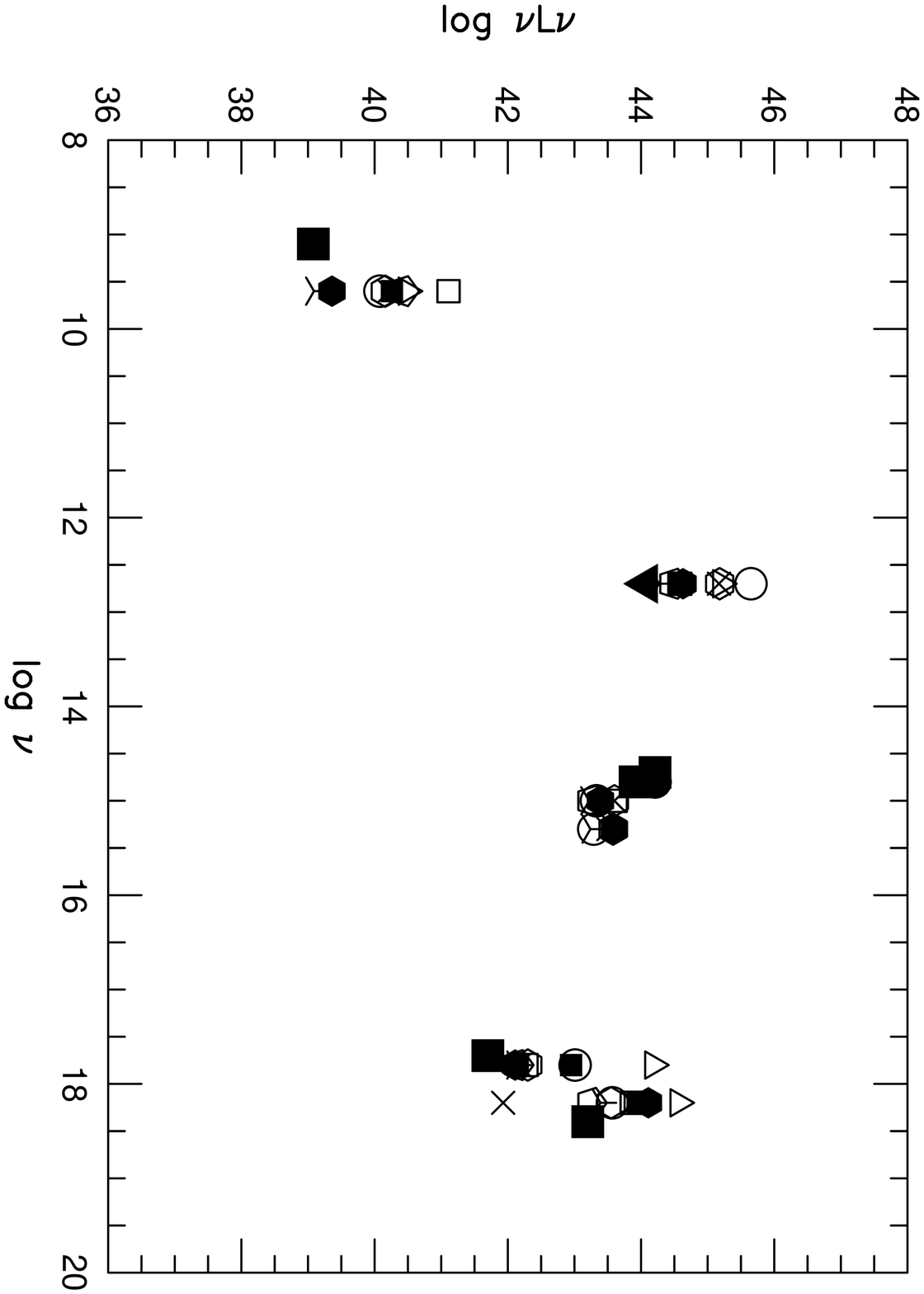]{
The spectral energy distribution of AX~J131501+3141.
Large filled rectangles are data points of AX~J131501+3141.
Large filled triangle is the upper limit to the infrared emission.
Other small marks represent SEDs of type~2 Seyferts
from Mas-Hesse et al.(1995):
open rectangles: Mrk~348,
crosses: NGC~1068,
filled rectangles: NGC~2110,
open pentagons: Mrk~3,
open hexagons: NGC~4388,
filled hexagons: NGC~4507,
open triangles: NGC~5506, 
arrows: NGC~5674,
open circles: NGC~7582.
 \label{fig3}}

\figcaption[]{{\it R} band image around AX~J131501+3141.
Galaxies~A, B, and C are marked. North is up and east is left.\label{fig4}}

\clearpage

\begin{table*}
\begin{center}
\begin{tabular}{cccccc} \hline
 band   &  name        &\multicolumn{2}{c}{coordinate (J2000)} & uncertainty & distance\tablenotemark{a}  \\ \hline
X-ray   & AX~J131501+3141 & $13^{\rm h} 15^{\rm m} 00.^{\rm s}9$ & $31^{\circ} 41' 28''$  & 30$''$    &  3.$''$2     \\
radio   &  FIRST source & $13^{\rm h} 15^{\rm m} 01.^{\rm s}19$ & $31^{\circ} 41' 29.''1$  & 1$''$    &  1.$''$1     \\
optical &  galaxy~A    & $13^{\rm h} 15^{\rm m} 01.^{\rm s}15$ & $31^{\circ} 41' 28.''1$  & 0.$''$8    &   ---     \\ \hline
\end{tabular}
\end{center}
\tablenotetext{a}{Distance from the optical center of galaxy A.}

\tablenum{1}
\caption{Coordinates of the X-ray, radio, and optical source. \label{tab1}}
\end{table*}

\clearpage

\begin{table*}
\begin{center}
\begin{tabular}{ccccccc}  \hline
No. & redshift &  {\it B}\tablenotemark{a}   & {\it R}\tablenotemark{a}   & log([NII]\tablenotemark{b}/H$\alpha$) & log([SII]\tablenotemark{c}/H$\alpha$) & log([OIII]\tablenotemark{d}/H$\beta$) \\ \hline  
galaxy~A   &  0.07203$\pm$0.00009 & 17.25 & 15.62  &  0.06\tablenotemark{e}            &     $-0.31$\tablenotemark{e}     &   0.65\tablenotemark{e}        \\ \hline
galaxy B   &  0.072   & 18.08 & 16.90  &  $-0.39$\tablenotemark{f}              &       -     & --         \\
galaxy C   &  0.189   & 18.75 & 17.03  &   0.0\tablenotemark{f}               &       -     & --         \\
\hline
\end{tabular}
\end{center}
\tablenotetext{a}{Total magnitude.}
\tablenotetext{b}{[NII]$\lambda$6583}
\tablenotetext{c}{[SII]$\lambda$6717+6731}
\tablenotetext{d}{[OIII]$\lambda$5007}
\tablenotetext{e}{Line ratio after subtraction of stellar continuum.}
\tablenotetext{f}{Line ratio from multi-slit spectroscopy without
subtraction of stellar continuum.}

\tablenum{2}
\caption{Results of the spectroscopic observations. \label{tab2}}

\end{table*}

\clearpage

\begin{table*}
\begin{center}
\begin{tabular}{ccccc} \hline                   
 name  &  band   & frequency log(Hz) & observed flux density         &      log($\nu L_{\nu}$)\tablenotemark{a}
 \\ \hline
FIRST   &  1.4 GHz  &   9.15 &     3.83 mJy           &     39.08 \\
{\it IRAS}    &  60$\mu$m &  12.70 &   $<$ 0.1 Jy           & $<$ 44.08 \\
optical &  {\it R} band   &  14.66 &    15.62 mag           &     44.21 \\
        &  {\it B} band   &  14.83 &    17.25 mag           &     43.91 \\
{\it ASCA} &  2.0 keV  &  17.68 & 9.6 $\times 10^{-15}$\tablenotemark{b} &     41.7 \\
           & 10.0 keV  &  18.38 & 6.4 $\times 10^{-14}$\tablenotemark{b} &     43.2 \\
\hline
\end{tabular}
\end{center}
\tablenotetext{a}{Unit in erg s$^{-1}$.}
\tablenotetext{b}{Flux density in erg s$^{-1}$ cm$^{-2}$ keV$^{-1}$ without absorption-correction.}

\tablenum{3}

\caption{Photometry of AX~J131501+3141. \label{tab3}} 

\end{table*}

\clearpage
\plotone{figure1.eps}

\plotone{figure2a.eps}

\plotone{figure2b.eps}

\plotone{figure3.eps}

\plotone{figure4.eps}

\end{document}